\documentclass[aps,prd,showpacs,superscriptaddress,nofootinbib,twocolumn,10pt]{revtex4-2}
\usepackage{color,amsmath,amssymb,graphicx,latexsym,lineno}
\usepackage{threeparttable,multirow,txfonts,subfigure,booktabs}
\usepackage[colorlinks=true,linkcolor=blue,citecolor=blue]{hyperref}
\nolinenumbers

\graphicspath{{./}{PRL_fig}}

\begin{document}

\title{Measurement of Very-high-energy Diffuse Gamma-ray Emissions from the Galactic Plane with LHAASO-WCDA}

\author{Zhen Cao}
\affiliation{Key Laboratory of Particle Astrophysics \& Experimental Physics Division \& Computing Center, Institute of High Energy Physics, Chinese Academy of Sciences, 100049 Beijing, China}
\affiliation{University of Chinese Academy of Sciences, 100049 Beijing, China}
\affiliation{TIANFU Cosmic Ray Research Center, Chengdu, Sichuan,  China}
 
\author{F. Aharonian}
\affiliation{Dublin Institute for Advanced Studies, 31 Fitzwilliam Place, 2 Dublin, Ireland }
\affiliation{Max-Planck-Institut for Nuclear Physics, P.O. Box 103980, 69029  Heidelberg, Germany}
 
\author{Axikegu}
\affiliation{School of Physical Science and Technology \&  School of Information Science and Technology, Southwest Jiaotong University, 610031 Chengdu, Sichuan, China}
 
\author{Y.X. Bai}
\affiliation{Key Laboratory of Particle Astrophysics \& Experimental Physics Division \& Computing Center, Institute of High Energy Physics, Chinese Academy of Sciences, 100049 Beijing, China}
\affiliation{TIANFU Cosmic Ray Research Center, Chengdu, Sichuan,  China}
 
\author{Y.W. Bao}
\affiliation{School of Astronomy and Space Science, Nanjing University, 210023 Nanjing, Jiangsu, China}
 
\author{D. Bastieri}
\affiliation{Center for Astrophysics, Guangzhou University, 510006 Guangzhou, Guangdong, China}
 
\author{X.J. Bi}
\affiliation{Key Laboratory of Particle Astrophysics \& Experimental Physics Division \& Computing Center, Institute of High Energy Physics, Chinese Academy of Sciences, 100049 Beijing, China}
\affiliation{University of Chinese Academy of Sciences, 100049 Beijing, China}
\affiliation{TIANFU Cosmic Ray Research Center, Chengdu, Sichuan,  China}
 
\author{Y.J. Bi}
\affiliation{Key Laboratory of Particle Astrophysics \& Experimental Physics Division \& Computing Center, Institute of High Energy Physics, Chinese Academy of Sciences, 100049 Beijing, China}
\affiliation{TIANFU Cosmic Ray Research Center, Chengdu, Sichuan,  China}
 
\author{W. Bian}
\affiliation{Tsung-Dao Lee Institute \& School of Physics and Astronomy, Shanghai Jiao Tong University, 200240 Shanghai, China}
 
\author{A.V. Bukevich}
\affiliation{Institute for Nuclear Research of Russian Academy of Sciences, 117312 Moscow, Russia}
 
% ...
\collaboration{The LHAASO Collaboration}


\nolinenumbers
\begin{abstract}\nolinenumbers
The diffuse Galactic gamma-ray emission is a very important tool used to study the propagation and
interaction of cosmic rays in the Milky Way. In this work, we report the measurements of the diffuse
emission from the Galactic plane, covering Galactic longitudes from $15^{\circ}$ to $235^{\circ}$ 
and latitudes from $-5^{\circ}$ to $+5^{\circ}$, in an energy range of 1 TeV to 25 TeV, with the 
Water Cherenkov Detector Array (WCDA) of the Large High Altitude Air Shower Observatory (LHAASO). 
After masking the sky regions of known sources, the diffuse emission is detected with $24.6\sigma$ 
and $9.1\sigma$ significance in the inner Galactic plane($15^{\circ}<l<125^{\circ}$, $|b|<5^{\circ}$) 
and outer Galactic plane ($125^{\circ}<l<235^{\circ}$, $|b|<5^{\circ}$), respectively. The WCDA 
spectra in both regions can be well described by a power-law function, with spectral indices of
$-2.67\pm0.05_{\rm stat}$ in the inner region and $-2.83\pm0.19_{\rm stat}$ in the outer region,
respectively. Combined with the Square Kilometer Array (KM2A) measurements at higher energies, 
a clear softening of the spectrum is found in the inner region, with change of spectral indices 
by $\sim0.5$ at a break energy around $30$ TeV. The fluxes of the diffuse emission are higher by 
a factor of $1.5-2.7$ than the model prediction assuming local CR spectra and the gas column 
density, which are consistent with those measured by the KM2A. Along Galactic longitude, the 
spatial distribution of the diffuse emission shows deviation from that of the gas column density. 
The spectral shape of the diffuse emission are possibly variation in different longitude region. 
The WCDA measurements bridge the gap between the low-energy measurements by space detectors and 
the ultra-high-energy observations by LHAASO-KM2A and other experiments.  These results 
suggest that improved modeling of the wide-band diffuse emission is required.
\end{abstract}

\pacs{95.85.Pw,98.70.Sa}

\maketitle

{\it Introduction.} --- One of the most fundamental unresolved problems in astrophysics is the 
origin and propagation of cosmic rays (CRs). Besides the direct measurements of energy spectra 
and arrival directions of individual composition of CRs, the diffuse $\gamma$-ray emission plays
a unique and complementary role in constraining the origin and propagation of CRs. The diffuse $\gamma$-ray 
emission is usually expected to be produced through interactions between CRs (nuclei and
electrons/positrons) and the interstellar medium (ISM) or radiation field
\cite{2000A&A...362..937A,2000ApJ...537..763S,2004ApJ...613..962S}. Modeling of the diffuse
emission in the high-energy $\gamma$-ray bands (HE, $\lesssim$0.1~TeV) is  consistent 
with the measurements at high and intermediate latitudes by Fermi Large Area Telescope
\cite{2012ApJ...750....3A}, supporting the basic framework of CR propagation and interaction 
in the Milky Way. 

\begin{figure*}[!htb]
\centering
\includegraphics[width=1.0\textwidth]{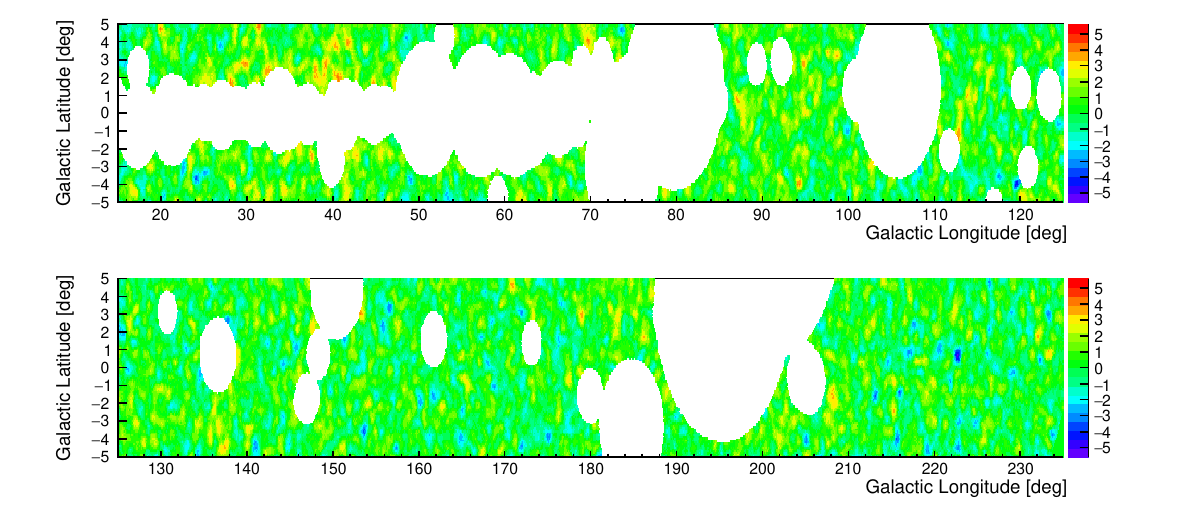}
\caption{The significance maps in Galactic coordinate of the inner Galaxy region (top panel) 
and outer Galaxy region (bottom panel)) for $N_{\rm hit}\geq200$ (corresponding to energies bigger 
than several TeV) after masking the resolved LHAASO and TeVCat sources.}
\label{fig:skymap}
\end{figure*}

Measurements of the wide-band diffuse $\gamma$-ray emission is particularly important 
in constraining the origin and testing the propagation models of CRs.
Due either to the small field-of-view of imaging atmospheric Cherenkov telescope arrays or
the limited sensitivity of extensive air shower detector arrays, the observations of diffuse 
$\gamma$-ray emission have been challenging for groundbased experiments for a long time.
A few measurements have been reported by MILAGRO
\cite{2007ApJ...658L..33A,2008ApJ...688.1078A}, H.E.S.S. \cite{2014PhRvD..90l2007A}, 
ARGO-YBJ \cite{2015ApJ...806...20B}, AS$\gamma$ \cite{2021PhRvL.126n1101A}, and HAWC
\cite{2024ApJ...961..104A}. Particularly, the recent precise measurement by the Square 
Kilometer Array (KM2A) of the Large High Altitude Air Shower Observatory (LHAASO) 
\cite{2023PhRvL.131o1001C} in the energy range from 10 to 1000 TeV, together with the Fermi 
measurement in space below TeV energy \cite{2023ApJ...957...43Z}, gives the diffuse emission 
in a very wide energy range and offers new insights in understanding the origin of the 
diffuse $\gamma$-rays and the propagation model for CRs (e.g.,
\cite{2024NatAs...8..628Y,2024PhRvD.109f3001Y,2023ApJ...957L...6F,2024PhRvD.109h3026D}).
Nevertheless, the detection of diffuse emission in energy range from 1 to 10 TeV is still 
lacking, which hinders a good discrimination of different models. In this work, we report the 
measurement of diffuse $\gamma$-ray emission from the Galactic plane in the energy range 
from 1 to 25 TeV using the Water Cherenkov Detector Array (WCDA) of the LHAASO experiment. 
The energy coverage of WCDA fills the gap between Fermi and LHAASO-KM2A. Together with Fermi 
and KM2A, the Galactic diffuse emission covering an energy range of more than 6 orders of 
magnitude (from sub-GeV to PeV) will be obtained for the first time.

{\it The LHAASO experiment.} ---
The LHAASO \cite{2022ChPhC..46c0001M} is a multi-component CR and $\gamma$-ray detection facility 
located on Haizi Mountain about 4410 meters above sea level in Daocheng, Sichuan province, China. 
It is designed to detect air showers induced by CR particles with energies from $\sim$TeV to 
$\sim$EeV and by photons from $\sim100$ GeV to $>$PeV. 
The WCDA is one of the three major components of LHAASO. It is a close-packed, surface water
Cherenkov detector facility with an area of 78,000 m$^2$. The WCDA consists of three ponds, 
two of them have an area of $150\times150$ m$^2$ with 900 detector units each, and the third
one has an area of $300\times110$ m$^2$ with 1320 detector units. The primary goal of WCDA is
to survey the northern $\gamma$-ray sky in the very-high-energy (VHE) band
\cite{2022ChPhC..46c0001M,2021ChPhC..45h5002A}. The large field-of-view and high sensitivity
of WCDA makes it a powerful detector for the study of diffuse $\gamma$-ray emission.

{\it Data analysis.} ---
The data used in this work is collected by WCDA between March 5, 2021 and July 31, 2023. 
After quality cuts, the total live time is 763 days. The event selection criteria are the 
same as in Ref.~\cite{2024ApJS..271...25C}. The parameter PINCness, defined as
${\mathcal P}=\frac{1}{N}\sum_{i=1}^N\frac{(\zeta_i-\langle\zeta_i\rangle)^2}{\sigma_{\zeta_i}^2}$,
where $\zeta_i$ is the logarithm of the $i$-th PMT charge, $\langle\zeta_i\rangle$ and 
$\sigma_{\zeta_i}$ are the average value and dispersion of a sample of gamma-like events, 
is applied to exclude hadronic showers \citep{2017ApJ...843...39A}. In this work we set
${\mathcal P}<1.1$. The region of interesting (ROI) is same as that used in the KM2A analysis \cite{2023PhRvL.131o1001C}, i.e., the inner Galaxy plane ($15^{\circ}<l<125^{\circ}$, 
$|b|<5^{\circ}$) and the outer Galactic plane ($125^{\circ}<l<235^{\circ}$, $|b|<5^{\circ}$. 
The analysis in different longitude regions are also performed. The events are binned by 
number of hits in six bins, i.e., 60-100, 100-200, 200-300, 300-500, 500-800, $\geq$800, 
which represent increasing energies from $\sim1$ TeV to $\sim25$ TeV. For the outer region, 
the measurement is performed for the last four bins to keep a relatively high significance. 
We also update the KM2A results for $25<E<1000$ TeV, following the analysis of 
\cite{2023PhRvL.131o1001C}, but using the full array data from July 20, 2021 to July 31, 
2023 and the updated source mask compared to the one used in this work.

The background, mainly from residual CR events, is estimated using the ``direct integral 
method'' \cite{2004ApJ...603..355F}. This method assumes that the detection efficiency is 
slowly varying and the time-average within a relatively short time window can properly 
reflect the background content. The time step to calculate the background is chosen to be 
1 hour, and the time window is chosen to be $\pm5$ hours of each step. Some spurious 
large-scale structures of the background may exist, which have been corrected in the 
analysis (see Fig. S1 of the {\tt Supplemental Material}) \cite{2023PhRvL.131o1001C}. 
Quite a number of pointlike and extended sources have been detected
\cite{2018A&A...612A...1H,2020ApJ...905...76A,2024ApJS..271...25C}. To reduce the impact of 
$\gamma$-rays sources on the background estimate, we mask the Galactic plane with latitudes
$|b|\leq10^{\circ}$ for declinations $\delta\leq 50^{\circ}$ and $|b|\leq5^{\circ}$ for 
$\delta>50^{\circ}$. The sky regions around sources out of the Galactic plane, based on 
the TeVCat as of December 2022~\cite{2008ICRC....3.1341W} and WCDA/KM2A source list \cite{2024ApJS..271...25C}, are also masked. The mask radius is chosen to be 5 times of
$\sigma\equiv\sqrt{\sigma_{\rm ext}^2+\sigma_{\rm psf}^2}$, where $\sigma_{\rm ext}$ is 
the fitted Gaussian extension of the source and $\sigma_{\rm psf}$ is the Gaussian width 
of the point spread function of WCDA. In this work we adopt $\sigma_{\rm psf}=0.5^{\circ}$ 
for all energy bins to keep the same sky regions. 

To measure the diffuse emission, we also mask known sources detected by 
LHAASO \cite{2024ApJS..271...25C} and those listed in TeVCat \cite{2008ICRC....3.1341W},
with different mask radii, to balance the remaining sky region and the impacts from sources. 
We adopt a mask radius of 2.5 times of $\sigma$ as defined above for each source.
For pulsar halos Geminga and Monogem, whose morphologies are not well described by simple 
Gaussian functions, the mask radius is set to be $8^{\circ}$ which can largely remove the 
emission from these two sources. The regions of interest (ROI) can be seen in 
Fig.~\ref{fig:skymap}, where blank regions are masked out. Since the source list and 
source extensions of WCDA are different from those of KM2A, the ROIs also differ slightly 
from the previous KM2A analysis \cite{2023PhRvL.131o1001C}. 
Specifically, the solid angle of the ROI is 0.172 (0.255) for the inner (outer) region in 
this work, compared with 0.206 (0.268) of the previous KM2A ROI \cite{2023PhRvL.131o1001C}.
We also estimate the residual contamination from resolved sources, through a fitting of 
the morphological distribution of the data for different rings surrounding resolved 
sources (represented by the multiplicative factor of source extension $\sigma$), using 
simulated templates of resolved sources and the diffuse emission. See Fig. S2 
of the {\tt Supplemental Material} for illustration. The contamination fraction, given in 
Table S1 of the {\tt Supplemental Material}, will be subtracted when 
evaluating the diffuse fluxes.

A likelihood fitting method is employed to derive the flux of the diffuse emission, in the 
whole energy band or individual energy bin. To properly account for the spectral shape of the
emission in a wide energy band, from TeV to PeV, a smoothly broken power-law (SBPL) function,
$\phi(E)=\phi_0 \left(E/10\,{\rm TeV}\right)^{-\alpha}
\left[1+(E/E_{\rm br})^{s}\right]^{(\alpha-\beta)/s}$,
is assumed, where $E_{\rm br}$ is the break energy, $\alpha$ and $\beta$ are the photon indices 
before and after the break, and $s$ characterizes the smoothness of the break which is fixed to 
be $5$ in this work. For the fitting of the WCDA data alone in a limited energy range, we assume 
a power-law spectrum $\phi(E)=\phi_0 \left(E/10\,{\rm TeV}\right)^{-\alpha}$ instead.
The test statistic (TS) is defined as two times of the logarithmic likelihood 
ratio of the signal plus background hypothesis ($H_1$) and the background only hypothesis 
($H_0$), i.e., ${\rm TS} = 2\ln({\mathcal L}_{s+b}/{\mathcal L}_b)$. The TS value follows 
approximately a $\chi^2$ distribution with two (four) degrees of freedom for PL (SBPL) model. 
The forward-folding method is used to determine the model parameters, i.e., for a given set 
of model parameters, we convolve the instrumental response functions to get the expected number 
of events, and then construct the Poisson likelihood according to the observations.

{\it Results on diffuse emission.} ---
Fig.~\ref{fig:skymap} shows the significance maps for $N_{\rm hit} \geq 200$ in Galactic coordinate 
for the inner (top panel) and outer (bottom panel) regions, respectively. The one-dimensional
significance distributions of these two regions and the control regions shifted by $20^{\circ}$ 
in Galactic latitudes are given in Fig. S3 of the {\tt Supplemental Material}.
The detection significance of diffuse emission is about $24.6\sigma$ ($9.1\sigma$) in the 
inner (outer) region. For the first time, the multi-TeV diffuse emission in the outer Galactic 
plane is detected by LHAASO-WCDA.

\begin{figure*}[!htb]
\centering
\includegraphics[width=0.48\textwidth]{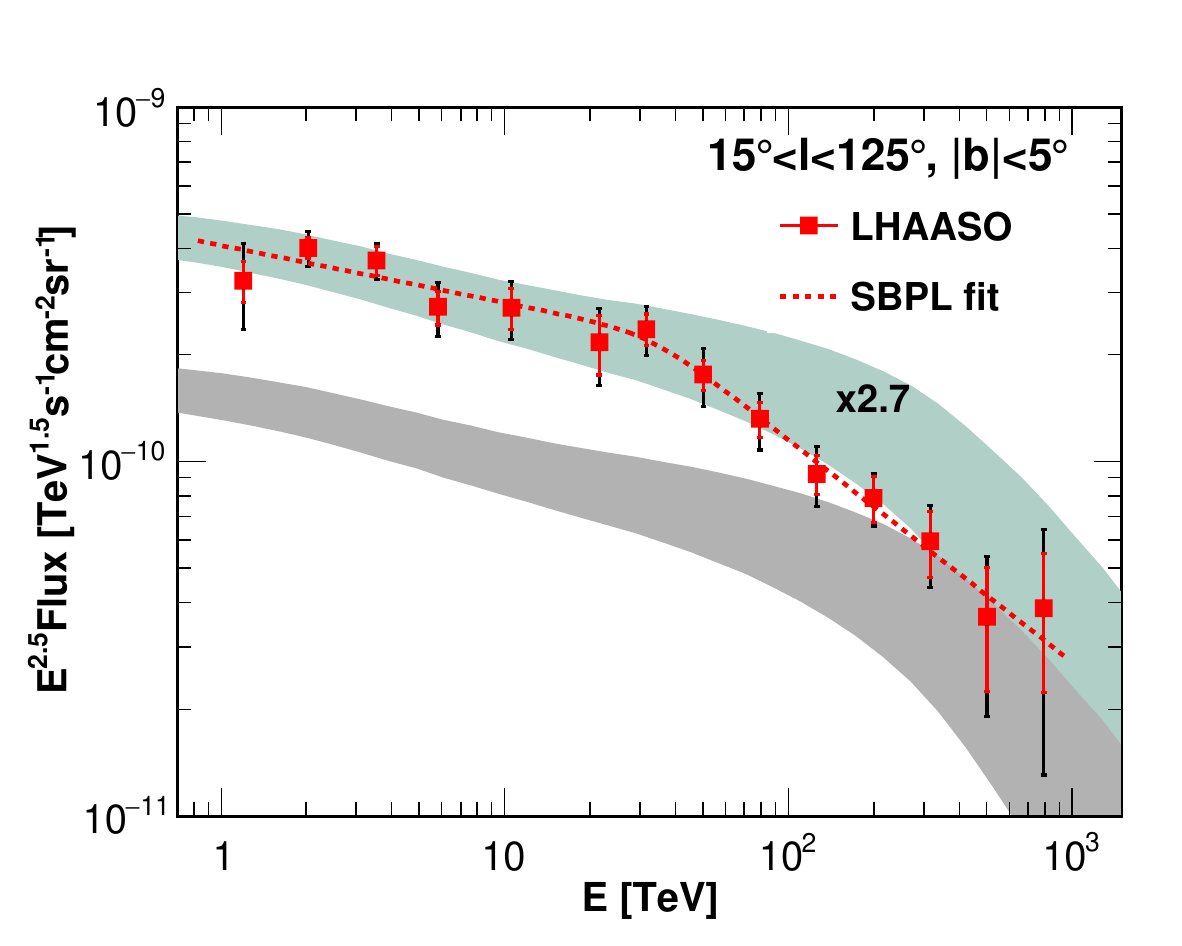}
\includegraphics[width=0.48\textwidth]{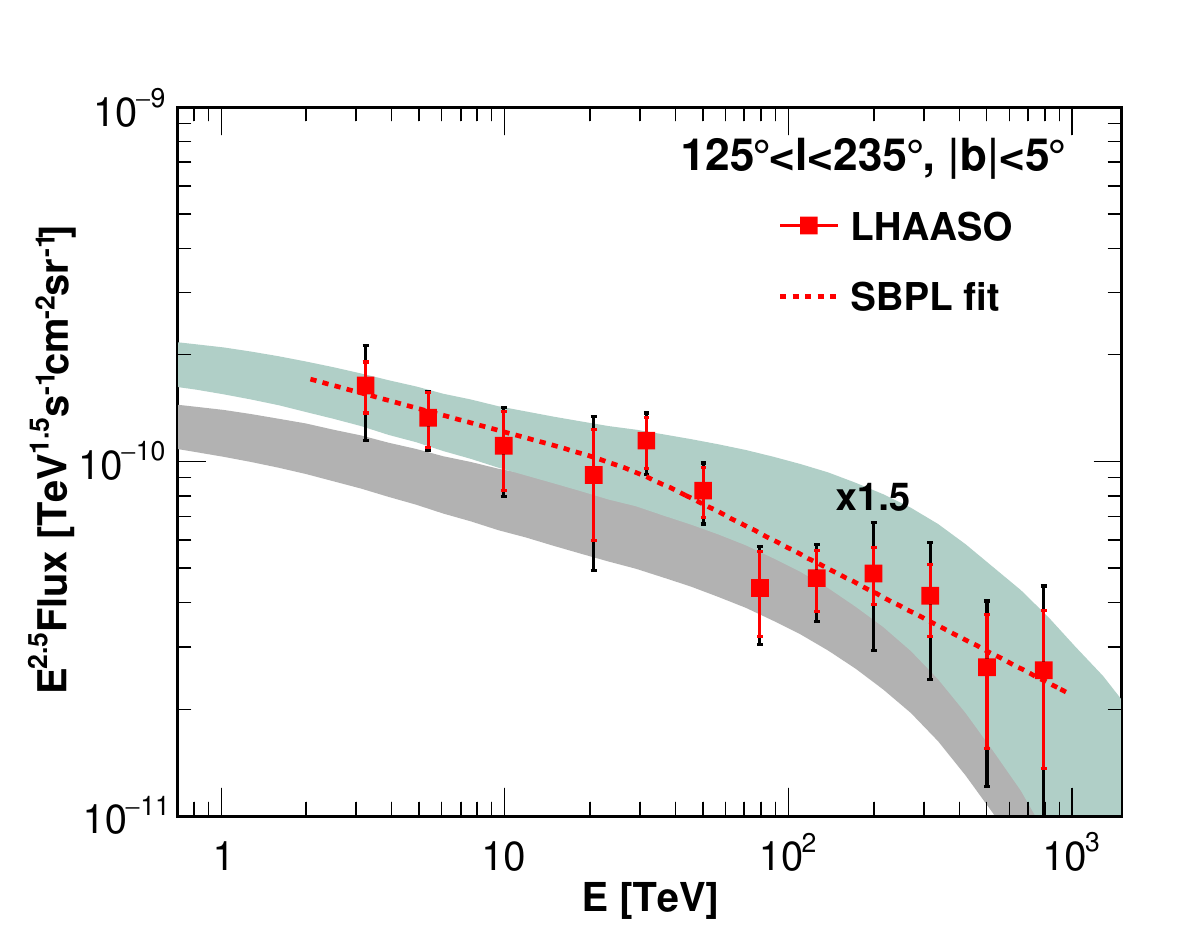}
\caption{SEDs of the diffuse $\gamma$-ray emission in the inner (left) and outer (right) regions.
Error bars are the statistical errors (short red) and the quadratic sum of the statistical and 
systematic errors (long black). Dashed lines show the best-fit SBPL spectra. The grey shaded
bands show the predicted diffuse fluxes from local CR intensities and the line-of-sight integral
gas content, and the cyan bands are upscaled by constant factors as labelled in the plots.}
\label{fig:spec}
\end{figure*}

The spectral energy distributions (SED) of the diffuse emission from $\sim$TeV to PeV in the two 
regions are given in Fig.~\ref{fig:spec}. The shorter error bars represent the statistical errors, 
and the longer ones are the quadratic sum of the statistical errors and the systematical 
uncertainties (see below for details). The data of the fluxes together with the statistical 
and systematic uncertainties can be found in Tables S2 and S3 of the {\tt Supplemental Material}. 
The diffuse $\gamma$-ray fluxes are lower by a factor of several than those of CR electrons and 
positrons \cite{2017Natur.552...63D,2019PhRvL.122j1101A,2023PhRvL.131s1001A,2024PhRvL.133v1001A}, 
as shown in Fig. S4 of the {\tt Supplemental Material}. However, in our data-driven background 
estimate method, the electron and positron background is largely suppressed given their nearly 
isotropic spatial distribution. For a better comparison with the neutrino emission
\cite{2023Sci...380.1338I}, we also present the total fluxes without source masks in Fig. S5 
of the {\tt Supplemental Material}.
 
\begin{figure}[!htb]
\centering
\includegraphics[width=0.48\textwidth]{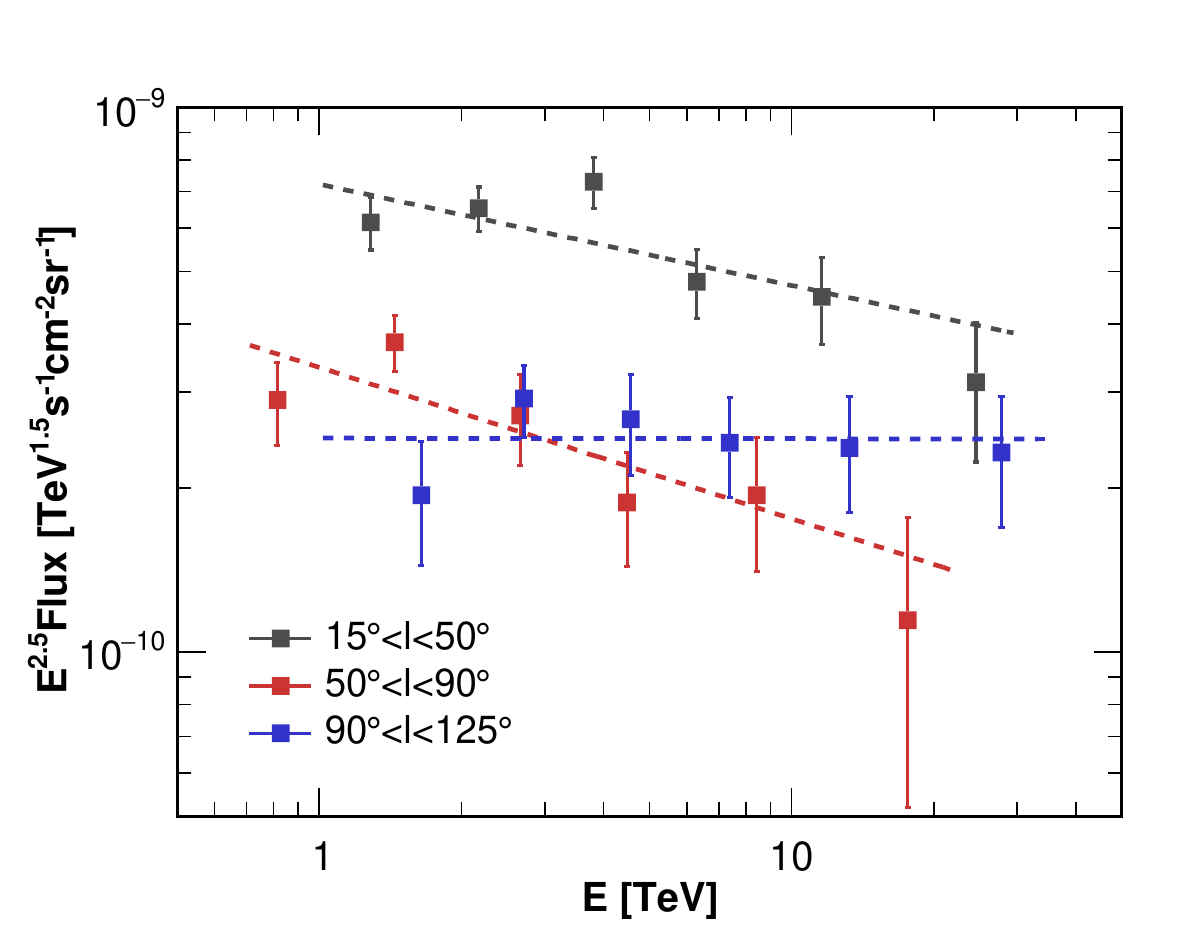}
\caption{Fluxes of diffuse $\gamma$-ray emission in the three sub-regions of the inner 
Galactic plane measured by the WCDA. Only the statistical errors are shown.}
\label{fig:specinner}
\end{figure}

The wide-band spectra are fitted with the SBPL function, with parameters being presented in 
Table \ref{table:fitspec_sbpl}. The fitting results indicate that breaks of the spectra, around 
$30$ TeV, exist. Quantitatively, a single power-law fitting gives a TS value of 
1099.10 (272.29) for the inner (outer) region, while the SBPL fitting gives 1131.25 (273.40). 
Therefore, the significance of the spectral break in the inner region is about $5.3\sigma$,
given two more free parameters. For the outer region the break is not statistically significant.
Within the statistical uncertainties, the spectral parameters in the inner and outer Galactic 
plane are consistent with each other. We further divide the inner region into three sub-regions, 
with Galactic longitudes of $l\in[15^\circ,50^\circ]$, $l\in[50^\circ,90^\circ]$ and
$l\in[90^\circ,125^\circ]$, and re-fit their spectra. The results are shown in 
Fig.~\ref{fig:specinner}. The spectral index of the third sub-region is found to be slightly 
harder ($\sim2\sigma$ significance) than the other two sub-regions, as can be seen in Table
\ref{table:fitspec_pl}. Such differences remain for the background estimate with different 
mask region ($|b| < 10^\circ$) and time window (24 hours). These results indicate that 
spectral variations may exist across the Galactic plane.

\begin{table*}[!htb]
\centering
\caption{SBPL fitting parameters of the wide-band diffuse emission measured by WCDA and KM2A}
\begin{tabular}{lccccc} 
\hline \hline
Region & $\phi_0$ at 10 TeV & $\alpha$ & $\beta$ & $E_{\rm br}$ \\ 
& ($10^{-13}~{\rm TeV^{-1}~cm^{-2}~s^{-1}~sr^{-1}}$) & & & (TeV) \\ 
\hline 
$15^{\circ}< l <125^{\circ}$ (inner) & $8.88 \pm 0.53_{\rm stat}$ & $-2.66 \pm 0.05_{\rm stat}$ & $-3.13 \pm 0.08_{\rm stat}$ & $32.84 \pm 11.16_{\rm stat}$ \\ 
$125^{\circ}< l <235^{\circ}$ (outer) & $3.84 \pm 0.37_{\rm stat}$ & $-2.72 \pm 0.10_{\rm stat}$ & $-2.92 \pm 0.10_{\rm stat}$ & $27.86 \pm 22.49_{\rm stat}$ \\ 
\hline \hline 
\end{tabular}
\label{table:fitspec_sbpl}
\end{table*}

\begin{table}[!htb]
\centering
\caption {Power-law fitting results of the LHAASO-WCDA diffuse emission.} 
\begin{tabular}{lccc}
\hline \hline
 Region & $\phi_0$ at 10 TeV & $\alpha$ \\
  & ($10^{-13}~{\rm TeV^{-1}~cm^{-2}~s^{-1}~sr^{-1}}$) &  \\
\hline
$15^{\circ}< l <125^{\circ}$ (inner) & $8.50\pm0.58_{\rm stat}$ & $-2.67\pm0.05_{\rm stat}$ \\
$125^{\circ}< l <235^{\circ}$ (outer) & $3.49\pm0.55_{\rm stat}$ & $-2.83\pm0.19_{\rm stat}$ \\
\hline
$ 15^{\circ}< l <50^{\circ}$ & $14.88\pm1.26_{\rm stat}$ & $-2.69\pm0.06_{\rm stat}$ \\
$ 50^{\circ}< l <90^{\circ}$ & $5.55\pm0.91_{\rm stat}$ & $-2.78\pm0.09_{\rm stat}$ \\
$ 90^{\circ}< l <125^{\circ}$ & $7.79\pm0.81_{\rm stat}$ & $-2.50\pm0.09_{\rm stat}$ \\
\hline
\hline
\end{tabular}
\label{table:fitspec_pl}
\end{table}

\begin{figure*}[!htb]
\centering
\includegraphics[width=0.33\textwidth]{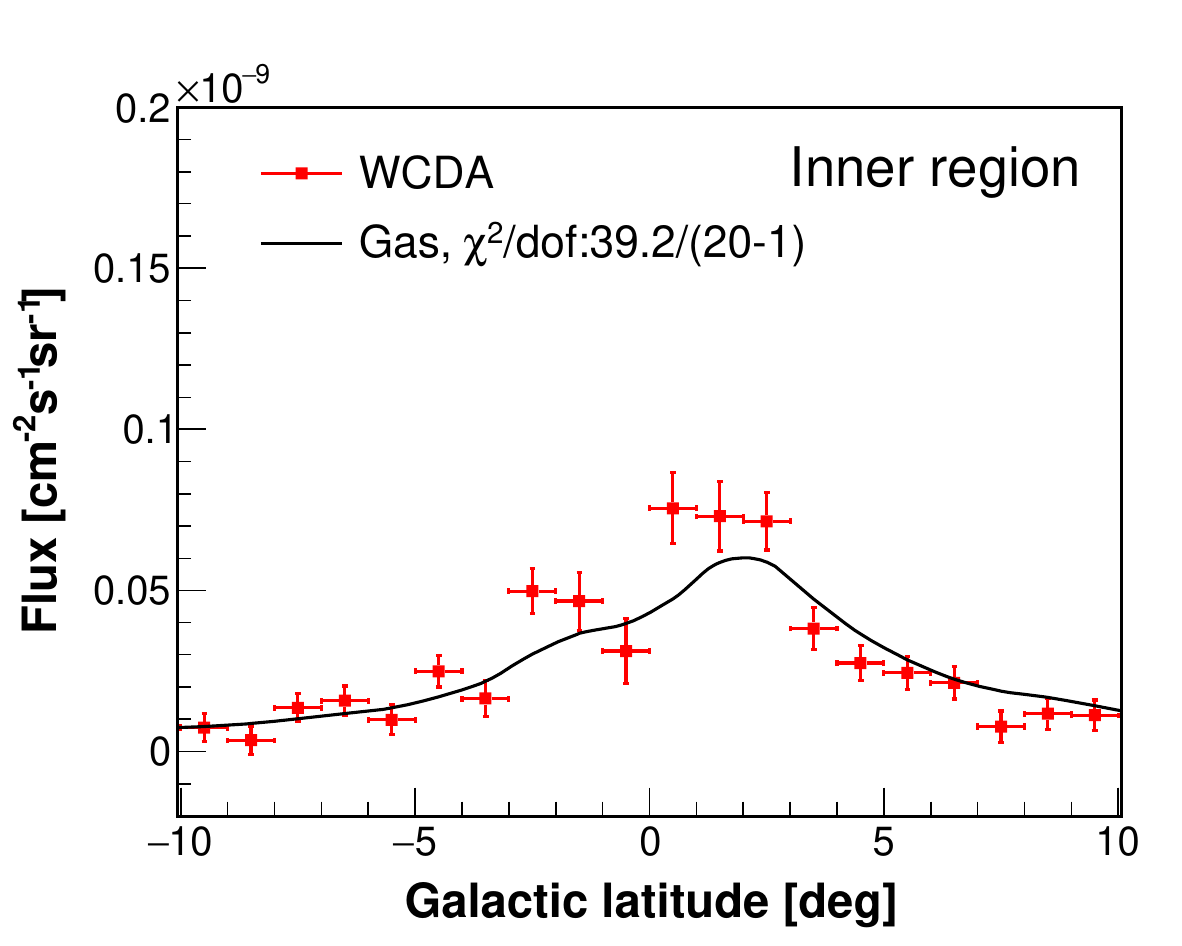}
\includegraphics[width=0.33\textwidth]{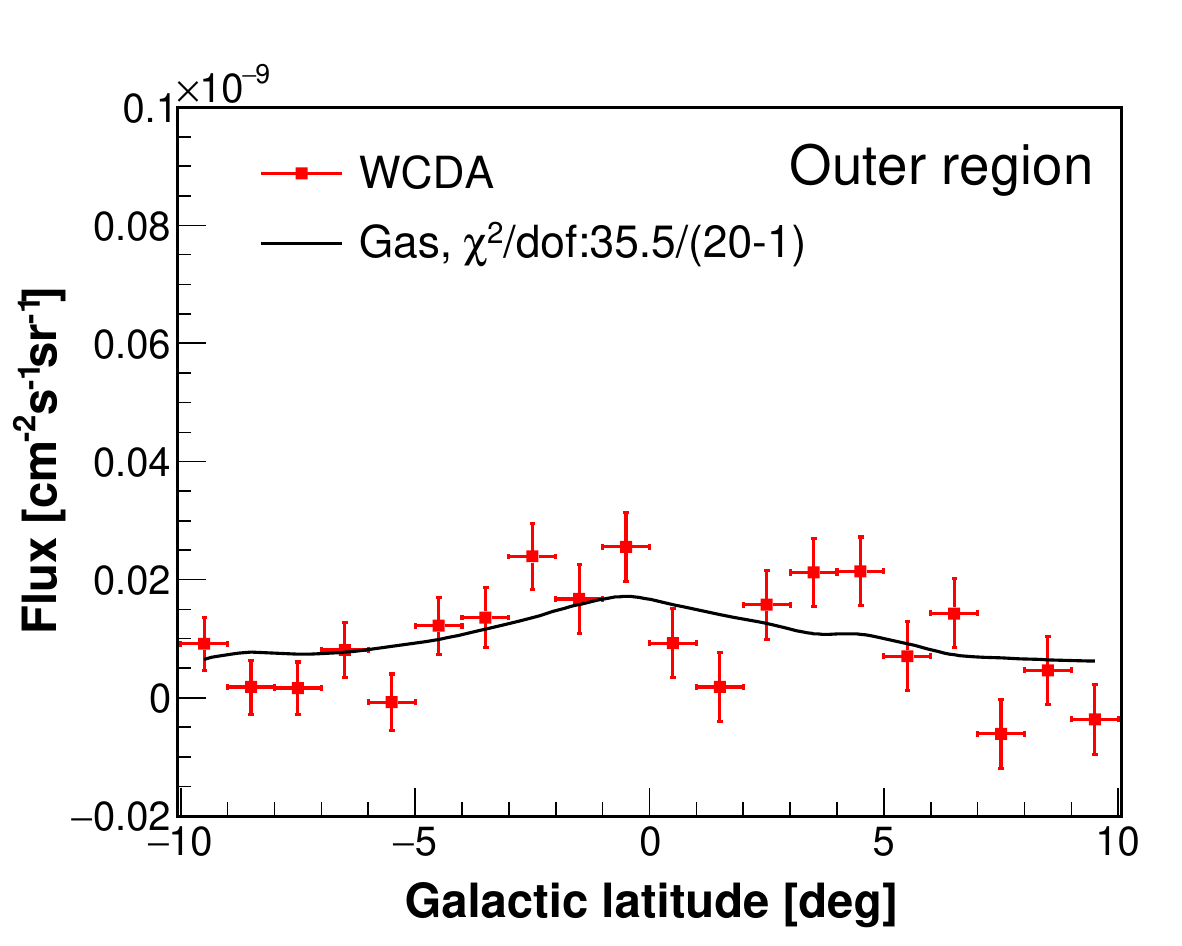}
\includegraphics[width=0.33\textwidth]{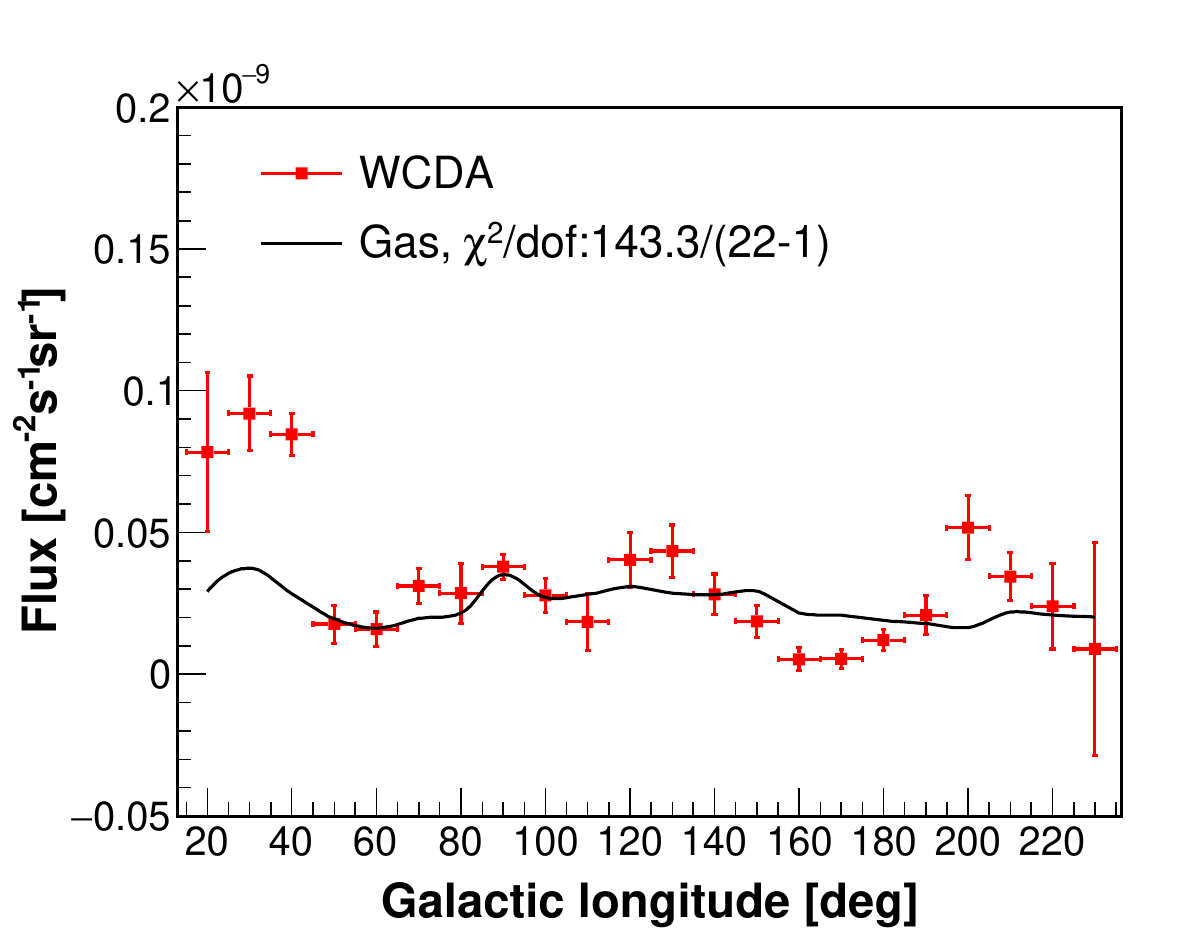}
\caption{Galactic latitude and longitude profiles of the diffuse fluxes for $N_{\rm hit}\geq 200$ 
(corresponding to an energy range of about $3-25$ TeV), compared with the gas 
distributions as shown by solid lines.}
\label{fig:1d_prof}
\end{figure*}

We also derive the spatial distributions of the WCDA diffuse emission along Galactic latitudes and
longitudes, as shown in Fig.~\ref{fig:1d_prof}. The results are compared with the gas distributions 
obtained from the sum of atomic hydrogen from the HI4PI survey \cite{2016A&A...594A.116H}, molecular
hydrogen from the CO survey \cite{2022ApJS..262....5D} with a CO-to-H$_2$ conversion factor of 
$X_{\rm CO}=1.6\times10^{20}$ cm$^2$ $({\rm K\,km\,s^{-1}})^{-1}$ \cite{2012ApJ...750....3A}, 
and an ionized hydrogen component from the model of \cite{1991Natur.354..121C}. 
An alternative way is to use the dust opacity to trace the gas column density
\cite{2016A&A...596A.109P}. Differences from these two methods vary for different regions 
of the Galaxy. On average, we find that the inferred gas column densities in the ROIs defined 
in this work are smaller than 10\%. 
It is shown that the latitude distributions are roughly consistent with the gas distributions 
in both the inner and outer regions. The fitting of the gas template to the data gives $p$-values 
of $0.003-0.013$ of the two regions. However, the longitude distribution deviates clearly from 
the gas distribution. The gas template fitting shows a $p$-value of $6.0\times10^{-21}$. 
Similar results have also been obtained for the KM2A measurements at higher energies
\cite{2023PhRvL.131o1001C} (see Fig. S6 of the {\tt Supplemental Material} for the comparison 
between WCDA and KM2A).

{\it Systematic uncertainties.} ---
One of the major sources of systematic uncertainties is the background estimate. We vary the
time window from $\pm5$ hours to $\pm12$ hours, and also change the mask regions for background 
estimate to $|b|\leq 10^{\circ}$ for all declinations to obtain the variations of the results.
The other effect related with the background is the spurious large-scale efficiency correction
(see Sec. A of the {\tt Supplemental Material}), for which we adopt different smoothing angles of 
$15^{\circ}$, $20^{\circ}$, $25^{\circ}$, and $30^{\circ}$. We find that the impacts of those 
changes on parameters $\phi_0$ are about $6\%~(16\%)$ and on $\alpha$ are about 0.05 (0.13), 
for the inner (outer) region. The variation of the atmosphere condition during the operation 
affects the detection efficiency, which is not properly modelled in the simulation. 
For point-like sources this effect is estimated to be about 8\% for the flux ($\phi_0$) and 
0.02 for the spectral index ($\alpha$) \cite{2024ApJS..271...25C}. Similar impacts on the 
diffuse emission is expected. Combining these results, we obtain the total systematic 
uncertainties as 10\% (18\%) for $\phi_0$ and 0.05 (0.13) for $\alpha$, for the inner (outer) 
region, respectively. The discussion of systematic uncertainties of the KM2A results can be 
referred to Ref.~\cite{2023PhRvL.131o1001C}. The systematic uncertainty of the flux in each 
energy bin has been shown in Fig.~\ref{fig:spec} and also in Tables S2 and S3 of the 
{\tt Supplemental Material}.

{\it Discussion.} ---
We compare the data with the predictions of the hadronic interactions between CRs and 
the ISM. We adopt the parameterizations of the local proton and helium spectra as given in
Ref.~\cite{2023PhRvL.131o1001C}, which include the uncertainties of the current measurements,
especially for those measured by indirect detection experiments above $\sim100$ TeV.
We assume that the CR intensity is uniform in the Galaxy\footnote{ Note that this 
assumption is over-simplified since variations of the CR intensities across the Galaxy 
were found to be present \cite{2016PhRvD..93l3007Y,2016ApJS..223...26A}. Assuming a
GALPROP model, which approximates the propagation halo as a cylinder, employs a numerical 
method to solve the transport equations, and includes experimental/observational inputs 
about the nuclear cross sections and gas distributions \cite{2000ApJ...537..763S}, we 
find that the expected diffuse fluxes are higher (lower) by 10\% (40\%) than the results
obtained here for the inner (outer) region after the same source mask.}, and calculate 
the $\gamma$-ray emission using the local CR spectra and the line-of-sight integrated 
gas content. The $\gamma$-ray production cross section used is from the {\tt AAfrag} 
package \cite{2019CoPhC.24506846K}. The expected diffuse $\gamma$-ray fluxes are shown 
by grey shaded bands in Fig.~\ref{fig:spec}. We can see that in the inner region the 
measurements are higher by a factor of $\sim2.7$ than the prediction, and in the outer 
region the measurements are higher by $\sim1.5$ times. These results are consistent with 
those found in the LHAASO-KM2A analysis \cite{2023PhRvL.131o1001C}. The HAWC measurements 
in a different sky region of the inner Galaxy also showed a factor of 2 higher fluxes 
compared with a DRAGON model prediction \cite{2024ApJ...961..104A}. 
In the outer region, the predicted spectrum is consistent with the data within 
uncertainties. We test the consistency of the spectral shape between the prediction and
the measurement in the inner region through a fit with free normalization of the model 
curve, and get $\chi^2=86.14$ and 19.16 for 13 degrees of freedom, for the ``high'' and 
``low'' predictions (corresponding to the upper and lower edges of the band shown in 
Fig.~\ref{fig:spec}). The $p$-values of the fittings are $7.61\times10^{-13}$ and 0.12, 
respectively. Given the current large uncertainties of the spectral measurements of 
individual CR species, we cannot draw a firm conclusion on whether the diffuse 
$\gamma$-ray data is consistent with the knee of the light CR components or not based 
on the spectral shape only.

\begin{figure*}[!htb]
\centering
\includegraphics[width=0.45\textwidth]{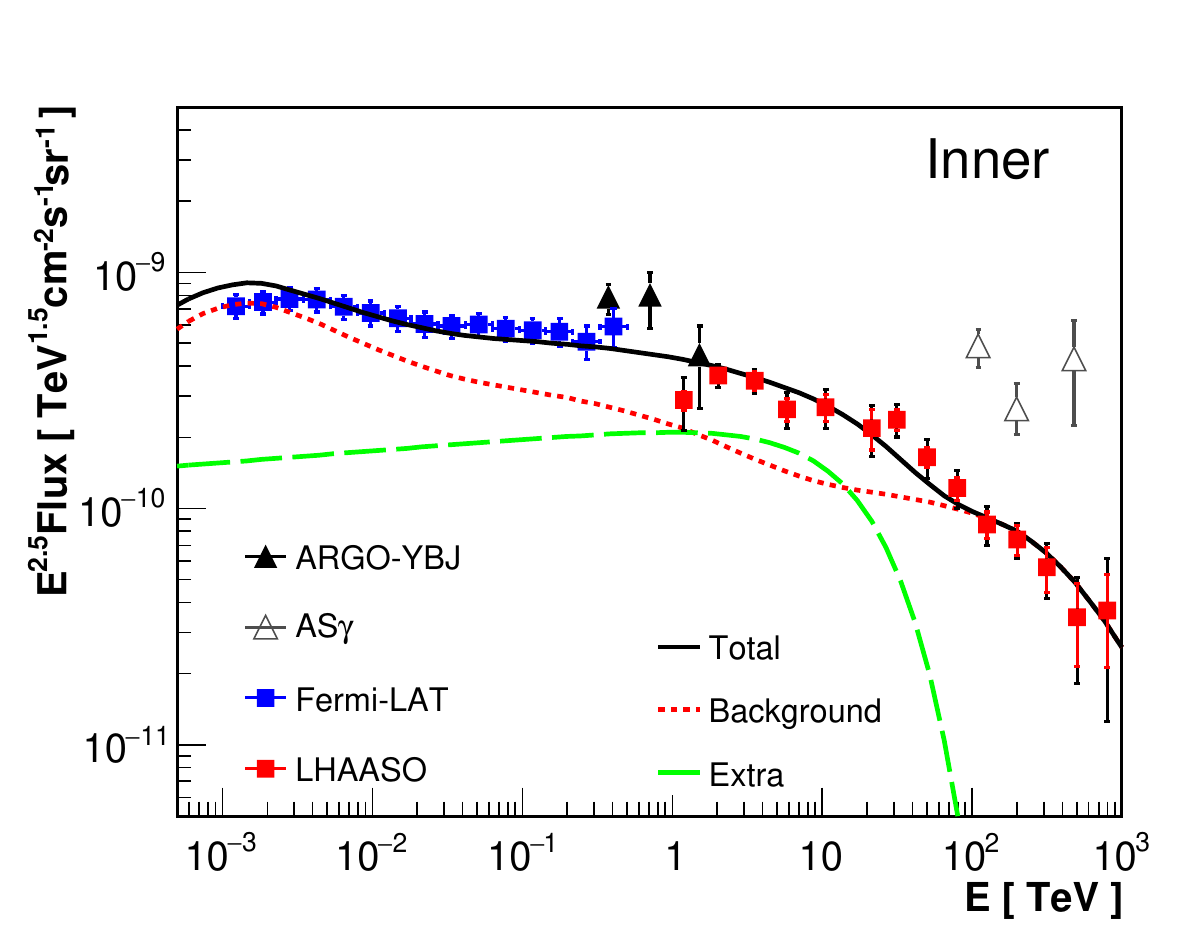}
\includegraphics[width=0.45\textwidth]{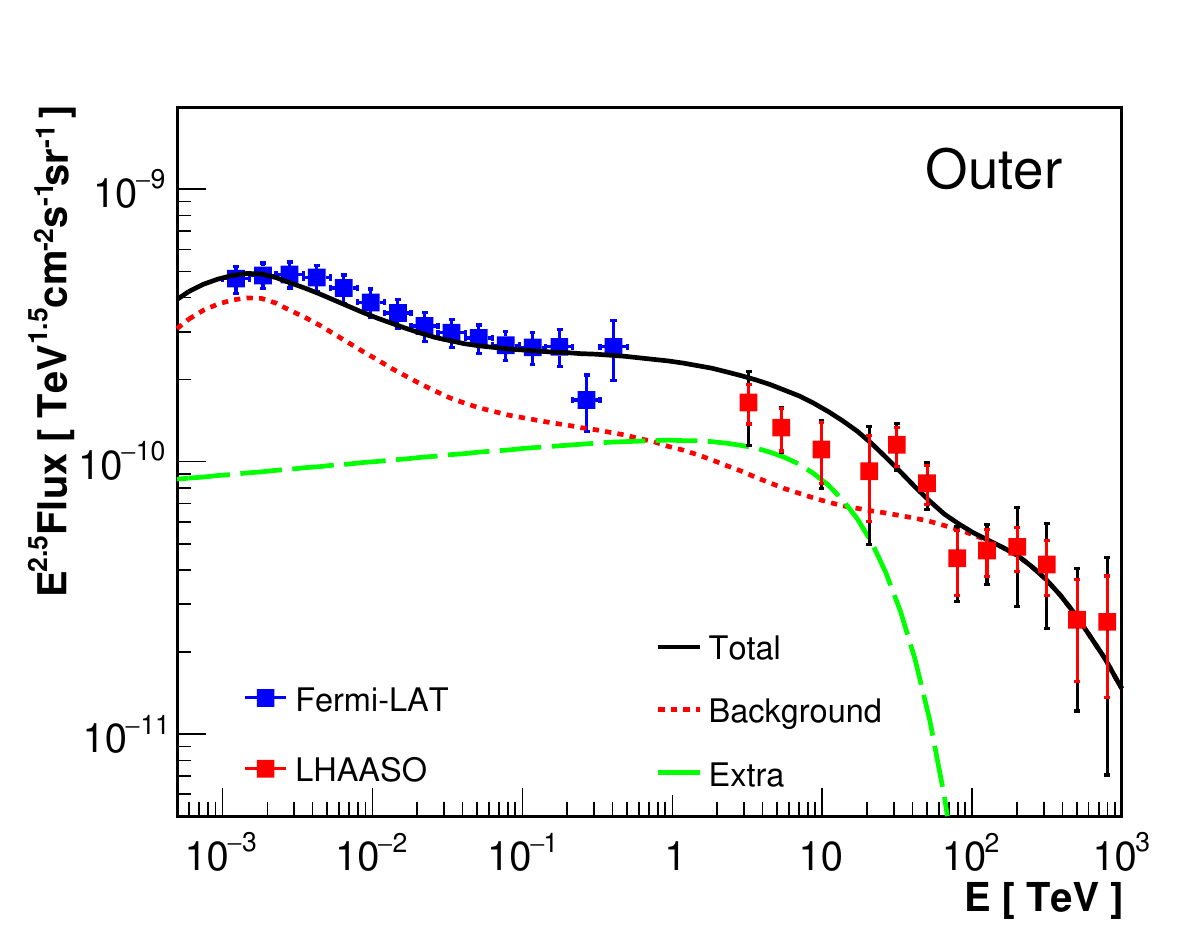}
\caption{Wide-band fluxes of diffuse $\gamma$-ray emission in the inner and outer Galaxy regions.
To make the comparison more directly, the Fermi-LAT results with the same mask used in this work. 
The ARGO-YBJ and AS$\gamma$ data in the left panel are for a slightly different region of
$25^\circ<l<100^\circ$ and $|b|<5^\circ$, with different source masks
\citep{2015ApJ...806...20B,2021PhRvL.126n1101A}. Red dotted lines show the prediction from 
a conventional CR propagation and interaction model \cite{2023ApJ...957...43Z}, green dashed 
lines show the contribution from an assumed extra component, and black solid lines are the 
sum of these two components.}
\label{fig:specall}
\end{figure*}

If we compare the data (including those measured by Fermi-LAT between 1 and 500 GeV
\cite{2023ApJ...957...43Z}, updated with the new mask of this work) with the GALPROP 
model prediction\footnote{ The propagation and source injection parameters were 
adjusted \cite{2020JCAP...11..027Y} to fit the up-to-date measurements of CR primary 
and secondary nuclei \cite{2017PhRvL.119y1101A,2018PhRvL.120b1101A,2016ApJ...831...18C}.
One should be aware that, a homogeneous, isotropic diffusion is assumed when calculating 
the CR propagation, which may be too simple \cite{2023arXiv230510251G}.} in a wider 
energy range, from GeV to PeV, we find that the excess mainly exists from several GeV 
to tens of TeV (see Fig.~\ref{fig:specall}). Such an excess may come from a population 
of unresolved, low surface brightness sources such as pulsar wind nebulae or pulsar halos
\cite{2000A&A...362..937A,2018PhRvL.120l1101L,2023Univ....9..381P,2024NatAs...8..628Y,
2024PhRvD.109h3026D}. The spatial variations of the spectra of the diffuse emission may 
further supports the hypothesis that unresolved sources may exist, although other 
possibilities such as spatially-dependent spectral slopes of CRs can not be excluded.
Following Ref.~\cite{2023ApJ...957...43Z}, we add an empirical component with 
spectrum $\propto E^{-2.45}\exp\left(-E/20\,{\rm TeV}\right)$ to the CR propagation model 
prediction, and find a good agreement with the wide-band measurements, as shown in 
Fig.~\ref{fig:specall}. Other sources like young massive star clusters 
\cite{2024arXiv240604087M}, the confinement and interaction of CRs around acceleration sources
\cite{2019PhRvD.100f3020Y,2022PhRvD.105b3002Z,2024ApJ...964...28H,2024arXiv241022199Y}, 
as well as the change of the canonical propagation scenario
\cite{2018PhRvD..97f3008G,2018PhRvD..98d3003L,2023A&A...672A..58D,2024arXiv240611015K} 
were also discussed in literature. The multi-messenger analysis of the diffuse $\gamma$ rays 
and neutrinos \cite{2023Sci...380.1338I} would be helpful in understanding the nature of the 
diffuse emission.

{\it Summary.} ---
In this work we measure the diffuse $\gamma$-ray emission in an energy range of about $1-25$ 
TeV from the Galactic plane using the LHAASO-WCDA data. The detection significance of the
diffuse emission, after masking sky regions around resolved sources, is $24.6\sigma$ for the
inner Galactic plane ($15^{\circ}<l<125^{\circ}$, $|b|<5^{\circ}$) and $9.1\sigma$ for the 
outer Galactic plane ($125^{\circ}<l<235^{\circ}$, $|b|<5^{\circ}$) regions, respectively. 
The diffuse emission from the outer Galactic plane is for the first time detected in the 
multi-TeV energy range. The WCDA measurements fill the gap between direct detection by space 
telescopes and ultra-high-energy measurements by groundbased experiments, and offer a full 
energy coverage from GeV to PeV for the Galactic diffuse $\gamma$-ray emission. The SEDs of 
the diffuse emission in the two regions have been measured with high precision. The WCDA 
spectra are consistent with power-law models, with indices of $-2.67\pm0.05_{\rm stat}$ and
$-2.83\pm0.19_{\rm stat}$. The joint WCDA-KM2A spectrum shows a break around 30 TeV in the 
inner region, with indices changing by about $0.5$. The break in the outer region is not 
significant. We also show that there are slight spectral variations across the Galactic 
plane in the inner region. The measured fluxes are higher by a factor of $1.5-2.7$ than 
the simple prediction of hadronic interactions between CRs (with locally measured fluxes) 
and the ISM, indicating that there are unmodelled source components or spatial variations 
of CR distribution.

{\it Acknowledgements.} ---
We would like to thank all staff members who work at the LHAASO site above 4400 meter above 
the sea level year-round to maintain the detector and keep the water recycling system, 
electricity power supply and other components of the experiment operating smoothly. We are 
grateful to Chengdu Management Committee of Tianfu New Area for the constant financial support 
for research with LHAASO data. We appreciate the computing and data service support provided 
by the National High Energy Physics Data Center for the data analysis in this paper. 
This work is supported by the following grants: the Department of Science and Technology 
of Sichuan Province, China (No. 2024ZYD0111, 24NSFSC2319), the National Natural Science Foundation of 
China (Nos. 12220101003, 12173039, 12393851, 12393852, 12393853, 12393854, 12205314, 12105301, 
12305120, 12261160362, 12105294, U1931201, 12375107, 12342502),  the Project for Young Scientists in Basic 
Research of Chinese Academy of Sciences (No. YSBR-061),  the National Science and Technology Development Agency (NSTDA) of Thailand, and the National Research Council of Thailand (NRCT) 
under the High-Potential Research Team Grant Program (N42A650868).

\bibliographystyle{apsrev}
\bibliography{paper_diffuse_wcda}

\end{document}